\documentclass[aps,prl,showpacs,twocolumn]{revtex4}
\usepackage{epsfig}
\usepackage{amsfonts}
\begin{document}

%%%%%%%%%%%%%%%%%%%%%%%%%%%%%%%%%%%%%%%%%%%%%%%%%%%%%%%%%%%%%%%%%%%%%
%%%%%%%%%%%%%%%%%%%%%         Title       %%%%%%%%%%%%%%%%%%%%%%%%%%%
%%%%%%%%%%%%%%%%%%%%%%%%%%%%%%%%%%%%%%%%%%%%%%%%%%%%%%%%%%%%%%%%%%%%%

\title{Influence of detector motion in Bell inequalities with entangled fermions}

%%%%%%%%%%%%%%%%%%%%%%%%%%%%%%%%%%%%%%%%%%%%%%%%%%%%%%%%%%%%%%%%%%%%%
%%%%%%%%%%%%%%%%%%%%     Authors & Addresses  %%%%%%%%%%%%%%%%%%%%%%%
%%%%%%%%%%%%%%%%%%%%%%%%%%%%%%%%%%%%%%%%%%%%%%%%%%%%%%%%%%%%%%%%%%%%%

\author{Andr\'e G. S. Landulfo and George E. A. Matsas}
\address{Instituto de F\'\i sica Te\'orica, Universidade Estadual
 Paulista,
         Rua Pamplona 145, 01405-900, S\~ao Paulo, S\~ao Paulo,
         Brazil}

\date{\today}

%%%%%%%%%%%%%%%%%%%%%%%%%%%%%%%%%%%%%%%%%%%%%%%%%%%%%%%%%%%%%%%%%%%%%%
%%%%%%%%%%%%%%%%%%%           Abstract            %%%%%%%%%%%%%%%%%%%%
%%%%%%%%%%%%%%%%%%%%%%%%%%%%%%%%%%%%%%%%%%%%%%%%%%%%%%%%%%%%%%%%%%%%%%

\begin{abstract}
We investigate how relativity influences the spin correlation of 
entangled fermions measured by moving detectors. In particular, 
we show that the Clauser-Horne-Shimony-Holt Bell inequality is 
not violated by quantum mechanics when the left and right 
spin detectors move fast enough.
\end{abstract}


\maketitle

%%%%%%%%%%%%%%%%%%%%%%%%%%%%%%%%%%%%%%%%%%%%%%%%%%%%%%%%%%%%%%%%%%%%%%%
%%%%%%%%%%%%%%%%%%%%%         Text Body          %%%%%%%%%%%%%%%%%%%%%%
%%%%%%%%%%%%%%%%%%%%%%%%%%%%%%%%%%%%%%%%%%%%%%%%%%%%%%%%%%%%%%%%%%%%%%%

The discovery of the Bell inequalities can be considered one of the
most important physics landmarks of the 20th century~\cite{B64}. It 
allows us to probe the essence of quantum theory by distinguishing it 
from local hidden variable theories. The genesis of this achievement 
can be traced back to the Einstein-Podolsky-Rosen discussion about the 
completeness of quantum mechanics~\cite{EPR35}. Presently the
 Clauser-Horne-Shimony-Holt~(CHSH) Bell inequality~\cite{CHSH69}  has been 
shown to be violated 30 standard deviations~\cite{WJSWZ98}, which
strongly supports quantum mechanics. In order to contribute to the
intense present debate on the interplay between relativity and 
quantum mechanics (see e.g. Refs.~\cite{SZGS02,PST02,PT02,TU03,PT04,KS05}), 
we investigate here how the former influences 
the spin correlation of entangled fermions measured by moving detectors. 
In particular we show that the CHSH Bell inequality can be satisfied 
rather than violated by quantum mechanics if the left and
right spin detectors are set in fast enough relativistic motion. We
adopt natural units $\hbar = c = 1$.

Let us assume a system composed of two spin-1/2 particles $A$ 
and $B$ with mass $m$ and zero total spin angular momentum. 
Each particle spin is measured along some arbitrary 
direction defined on the $ y \bot z $ plane. The distance between the planes 
along the $x$ axis is large enough to make both measurements causally disconnected. 
This is well known that local hidden variable theories satisfy the CHSH 
Bell inequality 
\begin{equation}
| E ({\bf a_2}, {\bf b_1}) + 
  E ({\bf a_2}, {\bf b_2}) + 
  E ({\bf a_1}, {\bf b_1}) -
  E ({\bf a_1}, {\bf b_2}) | 
\leq 2, 
\label{bellinequality}
\end{equation}
where
${\bf a_i}$  ($i=1,2$) are two arbitrary unit vectors contained in the 
$ y \bot z $ plane along which the spin ${\bf s}_A$ of particle $A$  
is measured, and analogously for the two arbitrary unit vectors ${\bf b_j}$ 
($j=1,2$) and spin ${\bf s}_B$ of particle $B$. 
Here 
\begin{equation} 
E ({\bf a_i}, {\bf b_j}) \equiv 
\lim_{N \to \infty} \frac{4}{N} \sum_{n=1}^N 
({\bf a_i} \cdot {\bf s}_A) ({\bf b_j} \cdot {\bf s}_B)
\label{correlation}
\end{equation}
is the spin correlation function obtained after an arbitrarily 
large number $N$ of experiments is performed, and 
${\bf a_i} \cdot {\bf s}_A $ and 
${\bf b_j} \cdot {\bf s}_B $ assume 
$\pm 1/2$ values.

Let us now test the inequality (\ref{bellinequality}) in the context of
quantum mechanics, where we allow the left and right detectors to move 
along the $x$ axis. For this purpose let us begin considering a 
quantum system composed  of two spin-1/2 particles. The corresponding 
normalized state can be written  as~\cite{H68, PST02} (see also 
Ref.~\cite{explicacao1})
\begin{equation}
|\psi\rangle = \sum_{s_A, s_B} 
               \int d \mathbf{p}_A d\mathbf{p}_B 
               \psi_{s_A s_B} (\mathbf{p}_A, \mathbf{p}_B)
               |s_A,  p_A \rangle 
               |s_B,  p_B \rangle, 
\label{est}
\end{equation}
where 
\begin{eqnarray}
\sum_{s_A, s_B} \int d\mathbf{p}_A d\mathbf{p}_B  
| \psi_{s_A s_B} (\mathbf{p}_A, \mathbf{p}_B) |^2 = 1, 
\\
\langle s'_X , p'_X | s_X , p_X \rangle = 
\delta_{s'_X \; s_X} \delta(\mathbf{p'}_X - \mathbf{p}_X),
\label{norm}
\end{eqnarray}
and $X=A, B$ distinguishes between both particles. Taking
$P_X^\mu$ and $\mathbf{S}_X \equiv {\bf s}_X \otimes I$
as the four-momentum and Wigner spin operators, respectively,
where ${\bf s}_X$ is half of the Pauli matrices, we have
\begin{eqnarray}
P_X^\mu |s_X,  p_X\rangle & = & p_X^\mu  |s_X,  p_X\rangle 
\nonumber \\
S_X^z  |s_X,  p_X\rangle & = & s_X  |s_X,  p_X\rangle
\nonumber
\end{eqnarray}
with 
$p_X = (\sqrt{\mathbf{p}_X^2 + m^2}, \mathbf{p}_X)$
and 
$s_X = \pm 1/2$.
Let us now assume that the two-particle system is prepared in 
a singlet state (see, e.g., Refs.~\cite{BLT75,PT04} for the two-spinor 
notation used below) 
\begin{eqnarray}
\psi(\mathbf{p}_A, \mathbf{p}_B)
&=&\frac{1}{\sqrt{2}}\biggl[ \left(
 \begin{array}{c} 
f_{\mathbf{k}_A}(\mathbf{p}_A)\\
0 \\  
\end{array} \right) \otimes \left( \begin{array}{c} 
0 \\
f_{\mathbf{k}_B}(\mathbf{p}_B)\\  
\end{array} \right) \nonumber \\
&-& \left( \begin{array}{c} 
0 \\
f_{\mathbf{k}_A}(\mathbf{p}_A)\\
\end{array} \right) \otimes  \left( \begin{array}{c} 
f_{\mathbf{k}_B}(\mathbf{p}_B)\\  
0 \\
\end{array} \right) \biggr]
\label{matrixsin}
\end{eqnarray}
from which we read 
\begin{eqnarray}
&& \psi_{s_A s_B} (\mathbf{p}_A,\mathbf{p}_B)
=
\frac{1}{\sqrt{2}} 
f_{\mathbf{k}_A} (\mathbf{p}_A) f_{\mathbf{k}_B}(\mathbf{p}_B) 
\nonumber \\
&& \times 
 (\delta_{s_A \ 1/2} \ \delta_{s_B \ -1/2} 
- \delta_{s_A \ -1/2} \ \delta_{s_B \ 1/2}).
\label{estsin}
\end{eqnarray}
We describe particles $A$ and $B$ by Gaussian packets:
$
f_{\mathbf{k}_X}(\mathbf{p}_X) = 
\pi^{-3/4}w^{-3/2}  
e^{-(\mathbf{p}_X - \mathbf{k}_X)^2/(2 w^2) }
$,
$w \in R_+$
and assume that they  move away 
from the origin in opposite directions at the same rate 
along the $x$ axis as defined in the laboratory frame: 
$\mathbf{k}_A = - \mathbf{k}_B = ( |k|, 0 , 0)$.

Now we are ready to discuss the spin measurement when the detectors acting on 
particles $A$ and $B$ have velocities ${\bf v}_{d_A}=(v_{d_A}, 0, 0)$  and ${\bf v}_{d_B}=(v_{d_B}, 0, 0)$, respectively. This is important to note that each detector 
will see  the wave function $|\psi \rangle$  in their proper frames transformed
through an unitary transformation ~\cite{H68, W96}:
\begin{equation}
|\psi \rangle \to |\psi'\rangle = 
U(\Lambda_{d_A})\otimes U( \Lambda_{d_B})|\psi \rangle,
\label{psi'}
\end{equation}
where
\begin{eqnarray}
&& U(\Lambda_{d_X} ) | s_X, p_X \rangle =
[ (\Lambda_{d_X} \ p_X)^0/p_X^0 ]^{{1}/{2}}  
\nonumber \\
&& \times \sum_{s'_X} D_{s'_X s_X} (\Lambda_{d_X}, p_X) 
| s'_X, \Lambda_{d_X} p_X \rangle.
\label{U}
\end{eqnarray} 
The Wigner rotation can be written in matrix form as
\begin{eqnarray}
D(\Lambda_{d_X}, p_X) = 
\frac{(p_X^0 + m) \sigma^0
\cosh (\alpha_{d_X}/2)}{[(p_X^0 + m)((\Lambda_{d_X} p_X)^0 + m)]^{1/2}}
\nonumber \\
+ 
\frac{ (p_X^x \sigma^0 + i \epsilon^{x i j} p_X^i \sigma^j ) 
\sinh (\alpha_{d_X}/2)}{[(p_X^0 + m)((\Lambda_{d_X} p_X)^0 + m)]^{1/2}},
\label{wigner}
\end{eqnarray}
where $\sigma^0$ and $\sigma^i$, $i=x,y,z$, are the 
usual $2 \times 2$ identity and  Pauli matrices,
respectively. We recall that
$$
\Lambda_{d_X}= 
\left( \begin{array}{cccc} 
\cosh \alpha_{d_X} & \sinh \alpha_{d_X} & 0 & 0\\
\sinh \alpha_{d_X} & \cosh \alpha_{d_X} & 0 & 0\\
0 & 0 & 1 & 0 \\
0 & 0 & 0 & 1\\  
\end{array} \right)
$$ 
with
$
\alpha_{d_X} \equiv - \tanh^{-1} v_{d_X}
$.
By using Eqs.~(\ref{est}) and~(\ref{U}) in Eq.~(\ref{psi'}), we obtain
\begin{eqnarray}
|\psi'\rangle & = & 
\sum_{s_X, s'_X} \int d\mathbf{p}_A  d\mathbf{p}_B  
\left(\frac{(\Lambda_{d_A} ^{-1} p_A)^0}{p_A^0}\right)^{\frac{1}{2}}
\left(\frac{(\Lambda_{d_B}^{-1} p_B)^0}{p_B^0}\right)^{\frac{1}{2}} 
\nonumber \\
&&\times
D_{s'_A  s_A} (\Lambda_{d_A}, \Lambda_{d_A}^{-1}p_A) 
D_{s'_B  s_B} (\Lambda_{d_B}, \Lambda_{d_B}^{-1}p_A) 
\nonumber \\
&&\times 
\psi_{s_A s_B} (\Lambda_1 ^{-1} \mathbf{p}_A, \Lambda_2 ^{-1} \mathbf{p}_B) 
|s'_A, p_A\rangle 
|s'_B, p_B\rangle, 
\label{transf}
\end{eqnarray}
where we have performed the change of variable 
$p_X \to \Lambda^{-1}_{d_X} p_X$ and we recall that 
$d {\bf p}_X/p^0 $ is a relativistic invariant. By
using Eqs.~(\ref{estsin}) and~(\ref{wigner}) in Eq.~(\ref{transf}), 
we can write $\psi'_{s_A s_B}(\mathbf{p}_A, \mathbf{p}_B)$, which 
appears in 
\begin{equation}
|\psi'\rangle = \sum_{s_A, s_B} 
               \int d \mathbf{p}_A d\mathbf{p}_B 
               \psi'_{s_A s_B} (\mathbf{p}_A, \mathbf{p}_B)
               |s_A,  p_A \rangle 
               |s_B,  p_B \rangle, 
\label{est'}
\end{equation}
using the two-spinor notation:
\begin{eqnarray}
\psi'(\mathbf{p}_A, \mathbf{p}_B)&=&\frac{1}{\sqrt{2}} \biggl[ \left(
 \begin{array}{c} 
a_1(\mathbf{p}_A)\\
a_2(\mathbf{p}_A) \\  
\end{array} \right) \otimes  \left( \begin{array}{c} 
b_1(\mathbf{p}_B)\\
b_2(\mathbf{p}_B) \\  
\end{array} \right) \nonumber \\
&-&
 \left( \begin{array}{c} 
-a_2(\mathbf{p}_A)\\
\overline{a_1}(\mathbf{p}_A) \\  
\end{array} \right) \otimes  \left( \begin{array}{c} 
\overline{b_2}(\mathbf{p}_B)\\
-b_1(\mathbf{p}_B) \\  
\end{array} \right) \biggr].
\end{eqnarray} 
This is the wave function on which the detectors will effectively
act to measure the particle spin. Here
\begin{eqnarray}
&& a_1({\bf p}_A) = 
K_A f_{{\bf k}_A}({\bf q}_A) [C_A (q_A^0 + m) + S_A (q_A^x + i q_A^y)], 
\nonumber \\
&& a_2({\bf p}_A)=
K_A f_{{\bf k}_{A}}({\bf q}_A)S_A q_A^z, 
\nonumber \\
&& b_1({\bf p}_B)=
-K_B f_{{\bf k}_{B}}({\bf q}_B)S_B q_B^z,  
\nonumber \\
&& b_2({\bf p}_B)=
K_B f_{{\bf k}_{B}}({\bf q}_B)[C_B (q_B^0 + m) + S_B (q_B^x - i q_B^y)], 
\nonumber
\end{eqnarray}
where
\begin{eqnarray}
K_X &\equiv& (q_X^0/p_X^0)^{1/2} /[(q_X^0 + m)(p_X^0 + m)]^{1/2},
\nonumber \\
q_X &\equiv& \Lambda_{d_X} ^{-1} p_X,
\nonumber \\
C_X &\equiv& \cosh(\alpha_{d_X}/2),
\nonumber \\
S_X &\equiv& \sinh(\alpha_{d_X}/2).
\nonumber 
\end{eqnarray}
%$K_X \equiv (q_X^0/p_X^0)^{1/2} /[(q_X^0 + m)(p_X^0 + m)]^{1/2}$,
%$q_X \equiv \Lambda_{d_X} ^{-1} p_X$,
%$C_X \equiv \cosh(\alpha_{d_X}/2)$
%and 
%$S_X \equiv \sinh(\alpha_{d_X}/2)$.
Next we trace out the momenta degrees of freedom since the detectors do only
measure spin. As a result, we obtain the following reduced density matrix:
\begin{eqnarray}
 \tau' \!
& = &\!
 \int d\mathbf{p}_A d\mathbf{p }_B 
 \psi'            (\mathbf{p}_A, \mathbf{p}_B)
{\psi'}^{\dagger} (\mathbf{p}_A, \mathbf{p}_B)
\nonumber \\
& = &\! 
(\rho_1 \otimes \rho'_1 \! -  \rho_2 \otimes \rho'_2  
 \! - \rho_3 \otimes \rho'_3  \! + \rho_4 \otimes \rho'_4)/2
\end{eqnarray}
with
\begin{eqnarray}
\rho_1 \otimes \rho'_1 & = & 
\left( \begin{array}{cc} 
1-V &  0 \\
0 & V \\  
\end{array} \right) \otimes  \left( \begin{array}{cc} 
W & 0 \\
0  & 1-W \\  
\end{array} \right),
\nonumber \\
\rho_2 \otimes \rho'_2  & = &
\left( \begin{array}{cc} 
0&  1-3V \\
-V & 0 \\  
\end{array} \right) \otimes  \left( \begin{array}{cc} 
0 & -W \\
1-3W & 0 \\  
\end{array} \right),
\nonumber \\
\rho_3 \otimes \rho'_3  & = & 
\left( \begin{array}{cc} 
0&  -V \\
1-3V & 0 \\  
\end{array} \right) \otimes  \left( \begin{array}{cc} 
0 & 1-3W \\
-W & 0 \\  
\end{array} \right), 
\nonumber \\
\rho_4 \otimes \rho'_4 & = &
\left( \begin{array}{cc} 
V &  0 \\
0 & 1-V \\  
\end{array} \right) \otimes  \left( \begin{array}{cc} 
1-W & 0 \\
0  & W \\  
\end{array} \right)
\nonumber
\end{eqnarray}
and
\begin{eqnarray}
V(\alpha_{d_A}) = 
\sinh^2 \left( \frac{\alpha_{d_A}}{2} \right)
\!\! \int \!\! d\mathbf{q}_A 
\frac{|f_{\mathbf{ k}_A}(\mathbf{q}_A)|^2 {q_A^z}^2 }{(q_A^0+m)(p_A^0+m)},
\label{V}
\\
W(\alpha_{d_B}) =
\sinh^2 \left( \frac{\alpha_{d_B}}{2} \right)
\! \!\int \!\! d\mathbf{q}_B
\frac{|f_{\mathbf{k}_B}(\mathbf{q}_B)|^2 {q_B^z}^2}{(q_B^0+m)(p_B^0+m)}, 
\label{W}
\end{eqnarray}
where we have used that 
$d\mathbf{p}_X/p_X^0= d\mathbf {q}_X/ q_X^0 $.
Now let us use our previous results to investigate 
Eq.~(\ref{bellinequality}). In quantum mechanical terms, the left-hand 
side of this equation can be expressed as
\begin{equation}
 |E ({\bf a_2}, {\bf b_1})\!  + \! 
  E ({\bf a_2}, {\bf b_2})\!  + \! 
  E ({\bf a_1}, {\bf b_1})\!  - \! 
  E ({\bf a_1}, {\bf b_2})| \!  
= \! |\langle C \rangle_{\tau'}| \! 
 \label{corrquant}
\end{equation}
where 
$\langle C \rangle_{\tau'} =  {\rm tr}( \tau' C )$ 
and
$$
C= 
  (\mathbf{\sigma \cdot a_2}) \otimes[\mathbf{\sigma \cdot (b_1 + b_2)}] 
+ (\mathbf{\sigma \cdot a_1}) \otimes[\mathbf{\sigma \cdot (b_1 - b_2)}].
$$ 
By using that
\begin{equation}
\langle  (\mathbf{\sigma \cdot u})\otimes(\mathbf{\sigma\cdot v})
\rangle_{\tau'} = -(1-2V)(1-2W)\; \mathbf{u \cdot v},
\end{equation}
where $ {\bf u}, {\bf v}  = {\bf a}_1, {\bf a}_2, {\bf b}_1, {\bf b}_2$, 
we cast  Eq.~(\ref{corrquant})  as
\begin{eqnarray}
|\langle C \rangle_{\tau'} | 
 = 
 |\langle  \mathbf{(\sigma \cdot a_2}) \otimes 
           \mathbf{(\sigma \cdot b_1})\rangle_{\tau'} + 
  \langle  \mathbf{(\sigma \cdot a_2}) \otimes 
           \mathbf{(\sigma \cdot b_2})\rangle_{\tau'} 
\nonumber \\ 
 + 
  \langle \mathbf{(\sigma \cdot a_1})\otimes 
          \mathbf{(\sigma \cdot b_1})\rangle_{\tau'} -  
  \langle \mathbf{(\sigma \cdot a_1})\otimes 
          \mathbf{(\sigma \cdot b_2})\rangle_{\tau'}|
\nonumber 
\end{eqnarray}
and finally as
\begin{equation}
|\langle C \rangle_{\tau'}|= (1-2V) (1-2W) \; |{\langle C \rangle_{\tau'}}^0 | 
\label{key}
\end{equation}
with 
$
|{\langle C \rangle_{\tau'}}^0 | = 
|\mathbf{a_2 \cdot b_1 + a_2 \cdot b_2} + \mathbf{ a_1\cdot b_1 - a_1\cdot b_2}|.
$
Eq.~(\ref{key}) is our key formula.
\begin{figure}
\epsfig{file=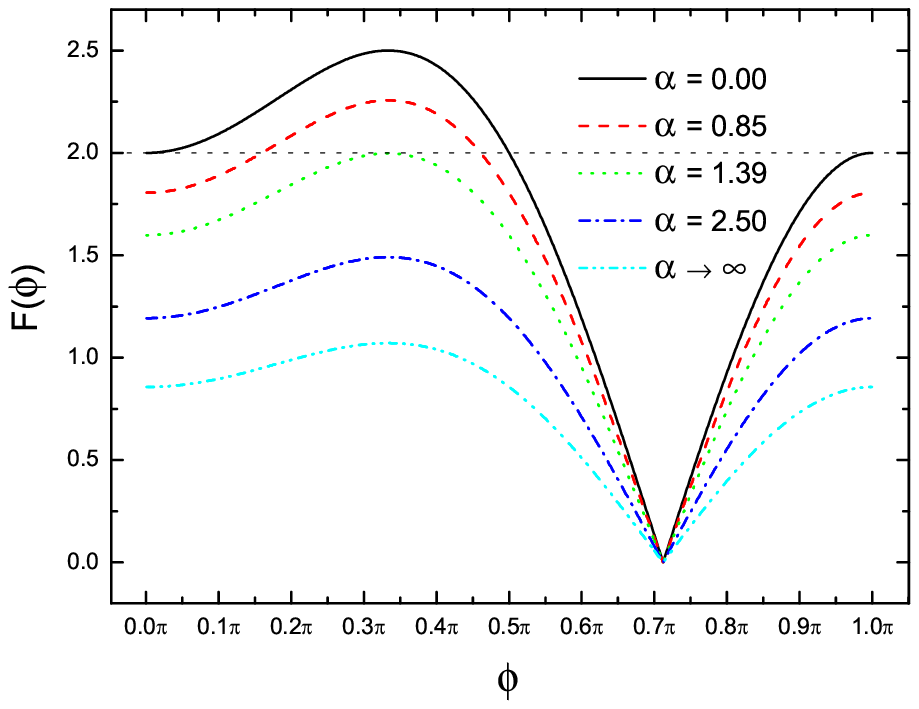,angle=0, width=8.5cm, height=6.cm}
\epsfig{file=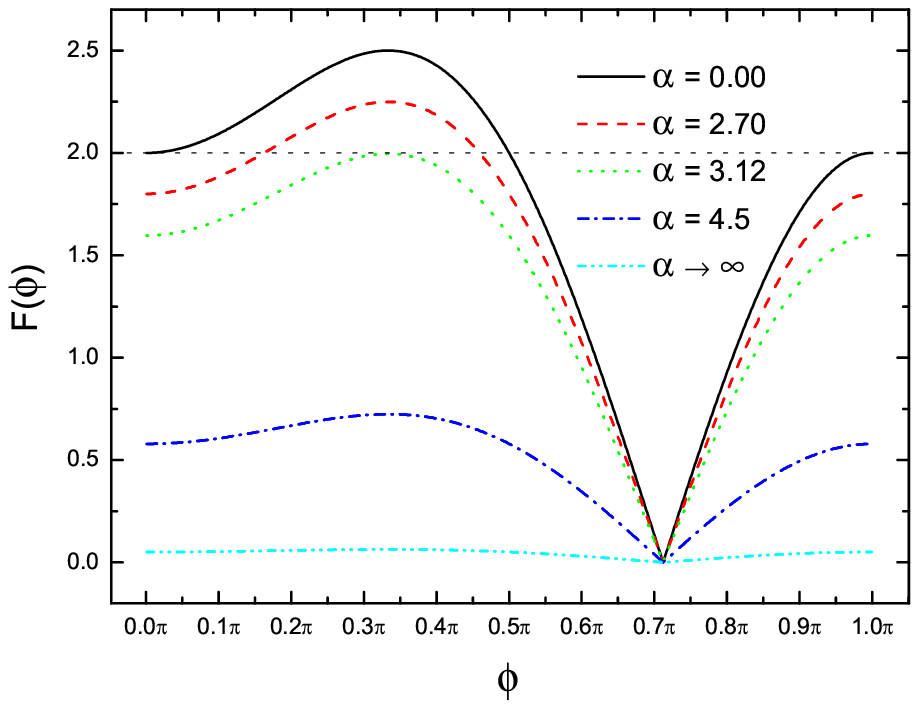,angle=0, width=8.5cm, height=6.cm}
\caption{ $F(\phi)$ as given in Eq.~(\ref{v}) is plotted as a
function of $\phi$ with $\tilde{w}=4$ for different values of 
$\alpha$. The top and bottom plots assume $|\tilde{k}|=0.01$  
and $|\tilde{k}|=100$, respectively. For $\alpha=0$ the usual
Bell inequality result is recovered. For $\alpha \gtrsim 1.39$, 
and $\alpha \gtrsim 3.12$, we have that $F(\phi) < 2$ for
the top and bottom plots, respectively.
\label{v1}}
\end{figure}
Note that when the detectors are at rest $\alpha_{d_A} = \alpha_{d_B}= 0$,
we recover the usual result: 
$
|\langle C \rangle_{\tau'}| =
|{\langle C \rangle_{\tau'}}^0 | 
$, 
i.e. the nontriviality introduced by the detector motion is isolated 
in the $(1-2V)(1-2W)$ multiplicative factor.
By defining $\alpha_i$, $\beta_i$ ($i= 1, 2$) as the angles between 
$\mathbf{a_i} $, $\mathbf{b_i}$ and the $x$ axis, respectively,
we get
\begin{eqnarray}
|{\langle C \rangle_{\tau'}}^0| 
& = & 
|\cos(\alpha_2 - \beta_1) + \cos(\alpha_2 - \beta_2)
\nonumber \\
& & +  \cos(\alpha_1 - \beta_1) - \cos(\alpha_1 - \beta_2) |.
\nonumber
\end{eqnarray}
For our purposes, this is sufficient to take the simpler case where
$\mathbf{a_2=b_1}$. By assuming this and  
$
\phi 
\equiv \cos^{-1} ( \mathbf{a_1 \cdot a_2} ) 
= \cos^{-1} ( \mathbf{b_1 \cdot b_2} ),
$
we obtain
$$
|\langle C \rangle_{\tau'}|_{\mathbf{a_2=b_1}} =
(1-2V)(1-2W)\; |1+ 2\cos\phi - \cos (2\phi)|. 
$$
\begin{figure}
\epsfig{file=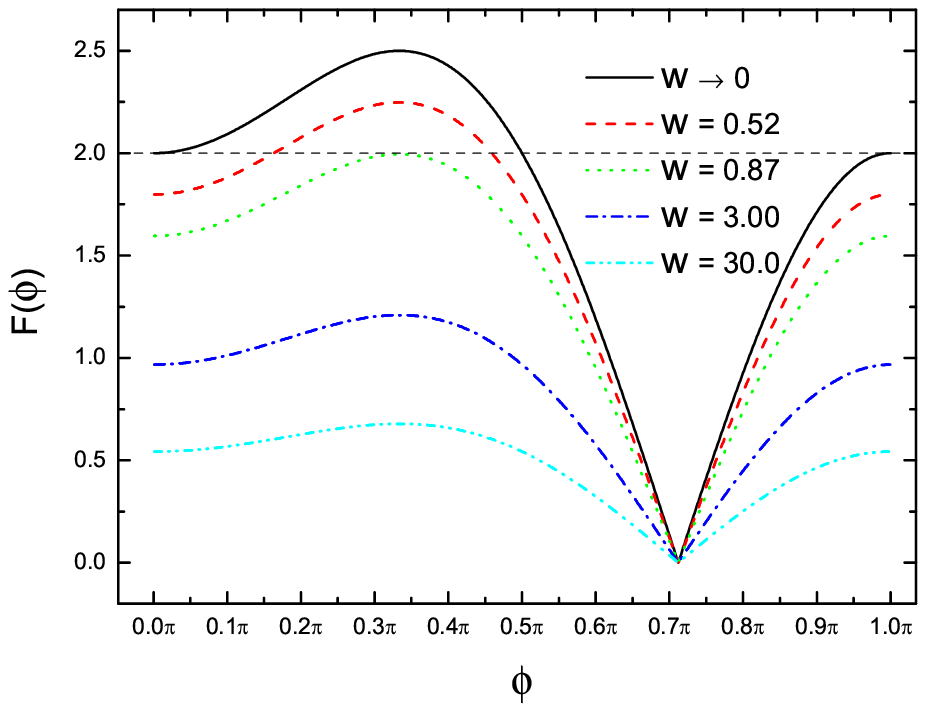,angle=0, width=8.5cm, height=6.cm}
\epsfig{file=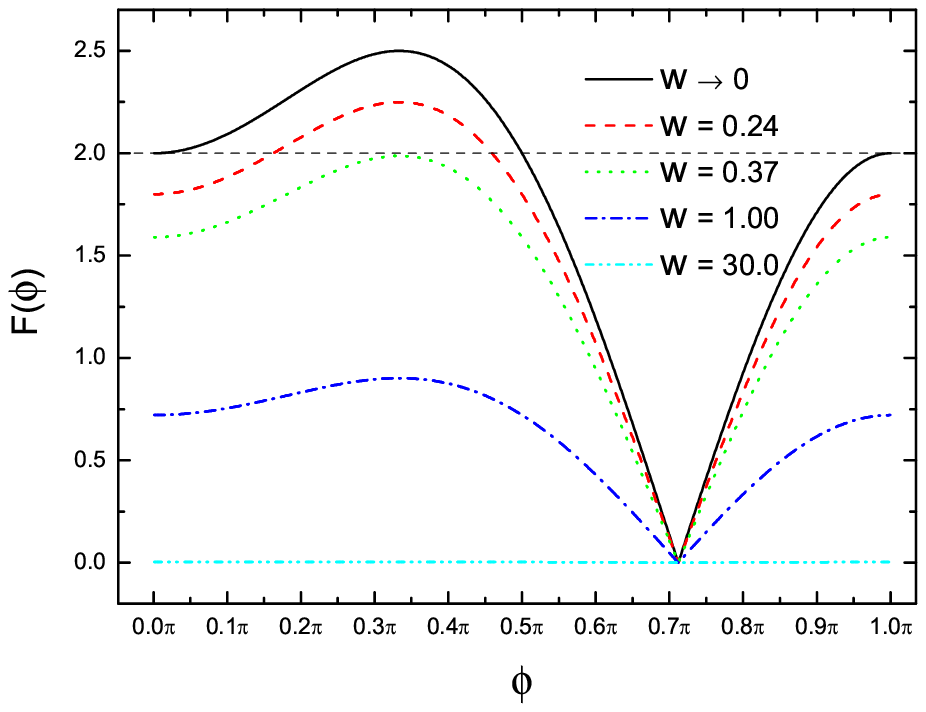,angle=0, width=8.5cm, height=6.cm}
\caption{ $F(\phi)$ as given in Eq.~(\ref{v}) is plotted as a
function of $\phi$ assuming $\alpha \to \infty$ for different values 
of the $\tilde w$ width. The top and bottom plots assume again 
$|\tilde{k}|=0.01$  and $|\tilde{k}|=100$, respectively. 
For $\tilde w = 0$ the usual Bell inequality result is recovered, 
while for $\tilde{w} \gtrsim 0.87$ and $\tilde{w} \gtrsim 0.37$
we have that $F(\phi ) < 2$ for the top and bottom plots, respectively.
\label{v6}}
\end{figure}
Let us, now, focus on the case where both detectors
boost away from each other with the same absolute 
rapidity~\cite{explicacao2}:
$ \alpha_{d_A}=-\alpha_{d_B}=-|\alpha| $, i.e. 
$v_{d_A} = -v_{d_B}=\tanh |\alpha| $. (For $|\tilde k| \ll 1$
we have verified that similar results are obtained no matter  
if the detectors move away or approach each other, as it 
should be.) For the sake of simplicity, we define
$$
F(\phi) \equiv 
|\langle C \rangle_{\tau'}|_\mathbf{a_2=b_1}^{\alpha_{d_A}=-\alpha_{d_B}}.
$$
Then, from Eqs.~(\ref{V})-(\ref{W}) we have
\begin{eqnarray}
V(-|\alpha|) 
&=& 
W(|\alpha|)
\nonumber \\
&=& 
\frac{\sinh^2 (|\alpha| /2) }{\sqrt{\pi} \tilde{w}^3} 
\!\! \int_{-\infty}^\infty \!\!\! d \tilde{q}^x 
\!\! \int_0^\infty \!\!\! d \tilde{q}^r  
G(\tilde{q}^x, \tilde{q}^r)
\label{int}
\end{eqnarray}
where we have used cylindrical coordinates with $q^x$ as the
symmetry axis and
\begin{equation} 
G(\tilde{q}^x, \tilde{q}^r ) 
=
\frac{(\tilde{q}^r)^3 \exp {[-((\tilde{q}^x-|\tilde{k}|)^2 +
(\tilde{q}^r)^2)/\tilde{w}^2}]} {(\tilde{q}^0 + 1) 
( \tilde{q}^0 \cosh |\alpha| - \tilde{q}^x \sinh|\alpha|  + 1) } 
\end{equation}
with 
$\tilde{q}^r=q^r/m$, 
$\tilde{q}^x=q^x/m$, 
$\tilde{q}^0 = \sqrt{ (\tilde{q}^x)^2 + (\tilde{q}^r)^2 +1}$, 
$|\tilde{k}|=|k|/m$
and
$\tilde{w}=w/m$. 
Then, we finally obtain the simple expression
\begin{equation}
F(\phi) =  F_0 |1+ 2\cos\phi - \cos (2\phi) |,
\label{v}
\end{equation}
where
$ F_0 = [1-2V(-|\alpha| )]^2$.
For very narrow wave packets in the momentum space, 
i.e. $\tilde{w} \ll 1$, this is easy to analytically solve 
the integral in Eq.~(\ref{int}) for particles moving slow 
enough, $\tilde{k} \approx 0$, and cast Eq.~(\ref{v}) as
\begin{equation}
F(\phi)|^{\tilde{k} \approx 0}_{\tilde{w} \ll 1} 
= \left( 1-\frac{\tilde{w}^2}{4} \tanh^2\frac{|\alpha|}{2}\right)^2 
  \!\! |1+ 2\cos\phi - \cos (2\phi) |.
\label{analitico}
\end{equation}
Clearly for $\tilde{w} \to 0$, we recover the standard Bell 
inequality result irrespective of the detector velocities, i.e. the 
nontriviality  driven by the detector motion in Eq. (\ref{analitico}) 
is not present when the entangled particles are described by 
momenta eigenstates~\cite{TU03,KS05}. This is so because 
only when particles are described by wave packets, $| \psi \rangle$ 
(which is a {\em pure state} according to observers lying at rest in 
the laboratory) looks like as a {\em mixed state} for the moving detectors 
once they ignore the momenta degrees of freedom~\cite{PST02}. (The 
corresponding ``missing information" gets hidden in the traced 
out momenta.) 

In Fig.~\ref{v1} we plot $F(\phi)$ for different detector velocities, i.e. 
$ |\alpha| $'s, assuming a  wave packet with 
$\tilde{w}=4$. The plots on the top and at the bottom take $|\tilde{k}|=0.01$ and $|\tilde{k}|=100$, respectively. We note that the standard Bell inequality 
result is recovered for $\alpha=0$ but this is not so when the detectors move.
Indeed, $F(\phi)$ decreases as the detector velocities increase. For 
$\alpha \gtrsim 1.39$, and $\alpha \gtrsim 3.12$, we have that $F(\phi) < 2$ 
in the whole $\phi$ range for the $|\tilde{k}|=0.01$ and $|\tilde{k}|=100$ 
cases, respectively, i.e. for these $\alpha$ intervals the Bell 
inequality is not violated for every $\phi$. Our numerical 
integration was cross checked against the Monte Carlo method and we have 
verified that it reproduces the analytic value given by Eq.~(\ref{analitico}) 
up to 1 part in $10^5$. Next we analyze how the packet width influences 
in our results. In Fig.~\ref{v6} we plot $F(\phi) $ for 
different $\tilde w $ values when the detectors have ultra relativistic
velocities: $\alpha \to \infty$. Again, we have assumed 
$|\tilde{k}|=0.01$ and $|\tilde{k}|=100$ for the top and bottom graphs, 
respectively. We see that
for $\tilde{w} \to 0$ we recover the usual Bell inequality 
result but as the wave packet width $\tilde w$ increases, $F(\phi)$ 
decreases. This reflects that the nontriviality associated with the 
detector motion is not manifest when the entangled particles are 
described by momenta eigenstates. For $\tilde{w} \gtrsim 0.87$ and 
$\tilde{w} \gtrsim 0.37$ we have that $F(\phi ) < 2$ in the whole 
$\phi$ range for the top and bottom plots, respectively.
Some experimental effort to verify the influence of the detector motion
in the Bell inequalities using photons can be found in the 
literature~\cite{SZGS02}. The natural generalization of our results 
for massless spin-1 particles would require the entanglement of 
horizontal/vertical polarized photons, which seems distinct from the one 
considered in Ref.~\cite{SZGS02}. This makes both results difficult to 
compare. Furthermore the replacement of fermions by photons is not quite
straightforward~\cite{PT02}. It would be interesting to verify our results 
in laboratory since this would also be an indirect test for all the 
underlying theoretical framework. Although conceptually the required 
experimental apparatus 
would be quite simple, this is not obvious to the present authors 
how difficult would be its realization in practice.

Modern physics is dominated by quantum mechanics and relativity.
This is fair to say that the Bell inequalities probe one of the deepest
aspects of quantum mechanics. Our analysis shows that the detector state of 
motion is crucial as one investigates the spin correlation of 
entangled fermions in the context of the Bell inequalities  
once one assumes the realistic physical situation where the particles of 
the entangled system are described by wave packets rather than by 
momentum eigenstates. 

\begin{acknowledgments}

We are grateful to A. O. Pereira for computational assistance.
A.L. and G.M. acknowledge full and partial support from Funda\c c\~ao
de  Amparo \`a Pesquisa do Estado de S\~ao Paulo, respectively. G.M. 
also acknowledges partial support from Conselho Nacional de
Desenvolvimento Cien\-t\'\i fico e Tecnol\'ogico.

\end{acknowledgments}


\begin{thebibliography}{99}
\bibitem{explicacao1} We are adopting Halpern~\cite{H68} and 
Peres~\cite{PST02} convention, where the relativistic factor 
$\sqrt{\mathbf{p}^2 + m^2}$ is introduced further in the Wigner
transformation law rather than in the normalization relation.
\bibitem{explicacao2} Similar conclusions are obtained when 
both detectors are boosted to the same direction.
\end{thebibliography}
\end{document}